\documentclass[useAMS, usenatbib]{aa}  
\usepackage{graphicx}
\usepackage{txfonts}
\usepackage{graphicx,lscape}
\usepackage{amssymb,amsmath}
\usepackage{rotating}
\usepackage{xcolor}

\def\msun{M_\odot}
\def\fs{f_{\mathrm S}}

\def\Izero{{I_{\mathrm 0}}}
\def\I0st{{I_{\mathrm 0}^{\mathrm{st}}}}
\def\V0{V_{\mathrm 0}}

\def\tE{t_{\mathrm E}}

\def\tEhelio{t_{\mathrm E}^{\mathrm{helio}}}
\def\te{t_{\mathrm E}}
\def\t0{t_{\mathrm 0}}

\def\u0{u_{\mathrm 0}}

\def\piE{\pi_{\mathrm{E}}}
\def\piEvec{\vec{\pi_{E}}}
\def\piEE{\pi_{\mathrm{EE}}}
\def\piEN{\pi_{\mathrm{EN}}}
\def\murel{\mu_{\mathrm{rel}}}
\def\murelvec{\vec{\mu_{\rm{rel}}}}
\def\muS{\mu_\mathrm{S}}
\def\muL{\mu_\mathrm{L}}

\def\thetaE{\theta_{\mathrm{E}}}
\def\pirel{\pi_{\mathrm{rel}}}
\def\piS{\pi_\mathrm{S}}
\def\piL{\pi_\mathrm{L}}
\def\DL{D_{\mathrm{L}}}
\def\DS{D_{\mathrm{S}}}
\def\ML{M_{\mathrm{L}}}

\newcommand{\eg}{{e.g.},}
\newcommand{\ie}{{i.e.},}

\begin{document} 

  \title{Constraining the masses of microlensing black holes and the mass gap with Gaia DR2}

\titlerunning{Microlensing BHs and mass gap with Gaia DR2}
\authorrunning{{\L}.~Wyrzykowski~\&~I.~Mandel}

   \author{
   {\L}ukasz Wyrzykowski\inst{1}\fnmsep\thanks{name pronunciation: {\it Woocash Vizhikovski}}
   \and
   Ilya Mandel\inst{2,3,4}
}          

\institute{
Warsaw University Astronomical Observatory, Al.~Ujazdowskie~4, 00-478~Warszawa, Poland, \email{lw@astrouw.edu.pl}
\and
Monash Centre for Astrophysics, School of Physics and Astronomy, Monash University, Clayton, Victoria 3800, Australia
\and
OzGrav: The ARC Center of Excellence for Gravitational Wave Discovery
\and 
Institute of Gravitational Wave Astronomy and School of Physics and Astronomy,
University of Birmingham, Edgbaston, Birmingham B15 2TT, United Kingdom
}

   \date{April 2019}

  \abstract  
  {  Gravitational microlensing is sensitive to compact-object lenses in the Milky Way, including white dwarfs, neutron stars, or black holes, and could potentially probe a wide range of stellar-remnant masses. 
However, the mass of the lens can be determined only in very limited cases, due to missing information on both source and lens distances and their proper motions. }
   {Our aim is to improve the mass estimates in the annual parallax microlensing events found in the eight years of OGLE-III observations towards the Galactic Bulge  
    with the use of Gaia Data Release 2 (DR2).}
   {We use Gaia DR2 data on distances and proper motions of non-blended sources and recompute the masses of lenses in parallax events. We also identify new events in that sample which are likely to have dark lenses; the total number of such events is now 18.}
   {The derived distribution of masses of dark lenses is consistent with a continuous distribution of stellar-remnant masses.  A mass gap between neutron star and black hole masses in the range between 2 and 5 solar masses is not favoured by our data, unless black holes receive natal kicks above 20-80 km/s. 
We present eight candidates for objects with masses within the putative mass gap, including a spectacular multi-peak parallax event with mass of $2.4^{+1.9}_{-1.3}\ M_\odot$ located just at 600 pc. The absence of an observational mass gap between neutron stars and black holes, or conversely the evidence of black hole natal kicks if a mass gap is assumed, can inform future supernova modelling efforts. 
}
   {}
 

   \maketitle
%

\section{Introduction}

Stellar mass black holes (BHs) are expected to be ubiquitous, with roughly one in a thousand stars ending their lives as a black hole  (\eg \citealt{ShapiroTeukolsky1983, Gould2000a, Lamberts2018}). However, they are elusive, and are generally easiest to find in interacting binaries.  For many years, Galactic BH X-ray binaries such as Cygnus X-1 provided the best studied sample of black holes.  The X-rays are emitted as a consequence of accretion of material from the companion star (either tidally stripped off by the black hole's gravity or ejected as stellar winds) onto the black hole (\eg \citealt{Shklovskii1967}, \citealt{Smak1982}, \citealt{Ziolkowski2010}, \citealt{2014Natur.514..202B}).  The variability in these X-rays imprinted by the binary's orbit  also provided the most accurate dynamical measurements of black hole masses (\eg \citealt{CasaresJonker2014}, \citealt{Corral-Santana2016}), although only about 20 accurate measurements are available.  The population of  black hole masses in these X-ray binaries extends upward from roughly $5 M_\odot$ \citep{Ozel2010,Farr:2010}.  This leads to a `mass gap' between the lightest known black holes and the heaviest known neutron stars (NSs) of approximately $2 M_\odot$ \citep{Demorest:2010,Antoniadis:2013, Swihart2017}.
To date, there has been only one discovery of a compact object in a non-interacting binary system with a giant star, with an uncertain compact-object mass measurement overlapping this mass gap \citep{Thompson2018}.

Recently, observations of gravitational waves  have roughly doubled the number of accurate black hole mass measurements.  During the first and second observing runs of the Advanced LIGO and Virgo detectors, ten binary black hole coalescences have been observed, with component masses ranging from roughly $7 M_\odot$ to more than $40 M_\odot$ \citep{BBH:O2}.  Moreover, a double neutron merger has been observed, GW170817, which left behind a compact object with a mass $\gtrsim 2.7 M_\odot$ \citep{GW170817}, which was  perhaps a supramassive neutron star that quickly collapsed into a black hole \citep{MargalitMetzger:2017};   regardless of its fate, it was the first confirmed single object in the putative mass gap.

However, the existing evidence for the mass gap should be treated with significant caution.  Both black hole X-ray binary observations and gravitational wave observations are sensitive to binaries in specific evolutionary phases, and require survival through a series of mass transfer events and supernova explosions \citep[e.g.,][]{FryerKalogera2001,TaurisvdH:2006,MandelFarmer:2018}.  Therefore, black hole masses measured in this way are subject to evolutionary selection biases, in addition to the more obvious detection biases that are easier to account for, such as the greater sensitivity of gravitational wave detectors to signals from higher mass systems.

Models of stellar evolution and collapse into neutron stars and black holes have a significant degree of uncertainty about the resulting mass distribution.  Some models (\eg \citealt{FryerKalogera2001, Kushnir2015}) did not find support for a mass gap, and others allowed for the possibility of either rapid explosions creating a mass gap or delayed explosions leading to a continuous mass distribution between NSs and BHs  (\eg \citealt{Fryer2012}).  However, recent models appear to find that minimum helium core masses for forming black holes of $\gtrsim 4 M_\odot$ \citep{Sukhbold:2016,Mueller:2016}; their models retain the entire helium core, so $4$ or $4.5 M_\odot$ becomes the minimum black hole mass.  On the other hand, it may be possible that that lower mass stars can form black holes with lower masses \citep{Couch:2019}, or that significant fall-back following a successful explosion would grow a lower mass seed into a black hole in the mass gap (though \citealt{Ertl:2016} do not find this in their longer simulations).  However, most of these are  one-dimensional simulations, as three-dimensional simulations are not yet mature enough to address the mass gap issue.  Furthermore, binary interactions could also play a role in the formation of the mass gap \citep{Belczynski2012}.

In addition to black holes of astrophysical origin, so-called primordial black holes may have formed from the collapse of overdensities in the very early Universe (\eg \citealt{Chapline1975}).  These have been conjectured as potential contributors to the dark matter content of the Universe.  Previous microlensing searches towards the Magellanic Clouds (\eg \citealt{Tisserand2007, Wyrzykowski2009}) have ruled out compact objects below 1 $\msun$ as a dominant component of the dark matter halo of the Milky Way (but see \citealt{Hawkins2015}).
In particular, OGLE-II and OGLE-III have monitored the LMC and SMC for about 13 years, and provided the toughest constraints on the dark matter halo fraction of about 4\% at 1 $\msun$ \citep{Wyrzykowski2011b}; however, their constraints for more massive lenses above 10 $\msun$ were much weaker, at about 20\%.  Therefore, there is still a window to explore and a theory to be tested that dark matter could be partially explained with more massive black holes of primordial origin (\eg \citealt{Clesse2015, Bird2016, Kashlinsky2016}). 

Additional observations of compact-object masses not subject to the biases present in X-ray binary and gravitational wave observations are needed to explore the existence of the mass gap and inform theoretical models.

        Gravitational microlensing provides an alternative tool for studying the distribution of masses of dark objects \citep{Paczynski1986, Griest1991, Paczynski1996}. This technique relies on continuous monitoring of millions of stars in order to spot a temporary brightening due to the space-time curvature caused by the presence and motion of a massive object, which could range from a planet to a black hole. Predictions based on the Galactic mass function and stellar synthesis \citep{Gould2000a} suggest that about 1\%\   of events should be caused by black holes, \ie there should be about 20 BH microlensing events happening every year. However, despite a 25 year history of photometric monitoring of hundreds of millions of stars by OGLE \citep{Udalski2015} and other surveys \citep[\eg][]{Alcock1997, Yock1998, Afonso2003}, only five candidates for lensing black holes have been proposed so far \citep{Mao2002,Bennett2002, Wyrz16}.

In order to obtain the mass of the lens \citep{Gould2000b}, it is necessary to measure both the angular Einstein radius of the lens ($\thetaE$) and the microlensing parallax ($\piE$):

\begin{equation}
M=\frac{\theta_\mathrm{E}}{\kappa \pi_\mathrm{E}},
\label{eq:mass}
\end{equation}where $\kappa=4G/(c^2 AU) = 8.144~\mathrm{mas/\msun}$; and $\piE$ is the length of the parallax vector $\piEvec$, defined as $\piE=\pirel/ \thetaE$, where $\pirel$ is relative parallax of the lens and the source. The microlensing parallax vector $\piEvec$ is measurable from the non-linear motion of the observer along the Earth's orbital plane around the Sun. The effect of microlensing parallax often causes subtle deviations and asymmetries relative to the standard Paczynski light curve in microlensing events lasting a few months or more, so that the Earth's orbital motion cannot be neglected. The parameter $\piEvec$ can also be obtained from simultaneous observations of the event from the ground and from a space observatory located $\sim$1AU away (\eg Spitzer or Kepler, \eg \citealt{Udalski2015Spitzer}, \citealt{CalchiNovati2015}, \citealt{Zhu2017K2Spitzer}). 

The angular size of the Einstein ring ($\thetaE$), on the other hand, is far more difficult to measure routinely in microlensing events. The Gaia space mission \citep{Gaia} collects time series of very precise astrometric measurements of positions of sources in microlensing events and therefore will enable measurements of  $\thetaE$ \citep{Rybicki2018, WyrzykowskiGaia16aye}, at least for the bright sub-sample of events happening concurrently with the Gaia mission in the years 2014-2020. 
The Einstein radius can also be obtained from the event timescale $\te$ if the relative proper motion $\murelvec$ of the lens and the source are known, since $\thetaE=\murel \te$ ($\murel$ is the length of the $\murelvec$ vector). To date, however, there have  only been a  limited number of   measurements of $\murelvec$ based on a detection of the luminous lens  a decade or so after the event \citep[\eg][]{KozlowskiHST, Batista2015, Beaulieu2018, Bramich2018, McGill2018, McGill2019}. 
Moreover, this method would not work for dark lenses, except for known pulsars, as shown by,  \citet{Dai2010}, \citet{Dai2015} and \citet{Ofek2018}, among others.


In \cite{Wyrz16} (hereafter  Wyrz16) we searched for long events observed over 8 years of the OGLE-III project (2001-2009) that exhibited a strong annual parallax signal. We obtained very broad mass functions for lenses by assuming that the lenses resided in the Galactic disk and followed the motion of the bulk of stars of the disk. Additionally, we restricted the dark lens candidate sample to events with source stars residing in the Red Clump region; therefore, we could assume a source distance of $\DS=8$ kpc. 
The relative proper motion was obtained by marginalising over a random selection of disk motion for lenses and bulge motion for sources.  This does not include the possibility that black holes could get significant natal kicks during supernovae (see Sect. 5 for a detailed discussion).

In April 2018 Gaia released its second data set, hereafter called GDR2, which contained five-parameter astrometric solutions for more than 1 billion Milky Way stars \citep{GaiaDR2}. The proper motions can now be directly used as measured by Gaia for sources which were lensed long before Gaia observed them, therefore without any perturbation to the astrometry by astrometric microlensing (\eg \citealt{Dominik2000}).

In this paper we reanalyse microlensing events with a strong parallax signal found in the OGLE-III data and with GDR2 five-parameter astrometric solutions in order to improve the constraints on the dark lens mass. The sample of events is described in Section 2 and lens mass distribution is derived in Section 3. In Section 4 we present the candidates for lenses with masses located in the mass gap. We discuss the results in Section 5 and conclude in Section 6.

\begin{table*} 
\centering
\caption{OGLE-III parallax events with Gaia Data Release 2 data and distances from CBJ18. 
}
\label{tab:gaia}
\begin{tabular}{lllllllll}
\hline
Lens name  &  GDR2 & parallax & $\mu_\mathrm{RA}$ & $\mu_\mathrm{Dec}$ & r$_\mathrm{est}$ & r$_\mathrm{lo}$ & r$_\mathrm{hi}$ & r$_\mathrm{len}$ \\
OGLE3-ULENS- &  sourceId & [mas] & [mas/yr] & [mas/yr] & [pc] & [pc] & [pc] & [pc]\\
\hline
\hline 
PAR-01 & 4050306756071281920 & 0.16$\pm$0.07 & -1.14$\pm$0.10 & -4.50$\pm$0.09 & 5148 & 3723 & 7774 & 2331\\
PAR-02$^*$ & 4062558472490201728 & 0.05$\pm$0.16 & -4.53$\pm$0.26 & -6.45$\pm$0.20 & 5883 & 3565 & 9917 & 2300\\
PAR-03$^*$ & 4063278480698674048 & 0.18$\pm$0.13 & 0.70$\pm$0.30 & -6.00$\pm$0.26 & 4526 & 2933 & 7643 & 2053\\
PAR-04$^*$ & 4068855482936627200 & 0.26$\pm$0.08 & -0.40$\pm$0.13 & -1.80$\pm$0.10 & 3516 & 2678 & 4985 & 1941\\
PAR-05$^*$ & 4041821408382645376 & 0.20$\pm$0.09 & -0.21$\pm$0.14 & -3.94$\pm$0.11 & 4406 & 3060 & 7072 & 2209\\
PAR-06 & 4062902035666302336 & 0.38$\pm$0.14 & 0.09$\pm$0.18 & -4.90$\pm$0.15 & 2756 & 1825 & 5045 & 2195\\
PAR-07$^*$ & 4056324241497959296 & 0.23$\pm$0.14 & -0.46$\pm$0.20 & -1.73$\pm$0.17 & 4138 & 2564 & 7634 & 2355\\
PAR-08 & 4041737712212920192 & -0.67$\pm$0.40 & -8.65$\pm$0.55 & -7.01$\pm$0.45 & 6146 & 3522 & 10304 & 2189\\
PAR-09$^*$ & 4043666182743831936 & -0.01$\pm$0.33 & -2.69$\pm$0.43 & -5.15$\pm$0.37 & 5446 & 2897 & 9791 & 2383\\
PAR-11 & 4050333105850811520 & 0.27$\pm$0.18 & -0.12$\pm$0.36 & -3.98$\pm$0.29 & 3893 & 2274 & 7563 & 2297\\
PAR-12$^*$ & 4053802962552499840 & 0.53$\pm$0.34 & 0.73$\pm$0.54 & -0.23$\pm$0.41 & 2901 & 1411 & 6629 & 2120\\
PAR-13$^*$ & 4061317643505236352 & 0.93$\pm$0.67 & -6.04$\pm$1.06 & -3.39$\pm$0.78 & 3661 & 1317 & 7887 & 2322\\
PAR-14 & 4056385337362836224 & 0.75$\pm$0.29 & -2.95$\pm$0.48 & -8.48$\pm$0.37 & 1654 & 945 & 4945 & 2369\\
PAR-15$^*$ & 4063554084450730880 & 0.28$\pm$0.25 & 2.53$\pm$0.62 & -0.65$\pm$0.48 & 3921 & 2097 & 7703 & 2175\\
PAR-16 & 4056068673902432000 & -0.51$\pm$0.17 & -2.57$\pm$0.33 & -7.94$\pm$0.26 & 9815 & 6503 & 14674 & 2388\\
PAR-17 & 4064872669535674240 & -0.09$\pm$0.55 & -0.68$\pm$1.19 & -3.40$\pm$0.99 & 4400 & 2224 & 8061 & 1982\\
PAR-19 & 4056570325872878720 & -1.05$\pm$0.30 & -1.49$\pm$0.40 & -6.47$\pm$0.35 & 8359 & 5251 & 13023 & 2345\\
PAR-20 & 4050327234484301824 & - & - & - & - & - & - & -\\
PAR-21 & 4041473382748323328 & 0.37$\pm$0.24 & -1.66$\pm$0.47 & -3.10$\pm$0.36 & 3260 & 1790 & 6891 & 2207\\
PAR-22$^*$ & 4043699752003532032 & 0.36$\pm$0.39 & -3.40$\pm$0.74 & -2.45$\pm$0.64 & 4228 & 1954 & 8515 & 2395\\
PAR-23 & 4063098817948361728 & -0.31$\pm$0.14 & 0.31$\pm$0.24 & -1.59$\pm$0.20 & 9031 & 6059 & 13414 & 2178\\
PAR-24$^*$ & 4063010749598773248 & -0.05$\pm$0.16 & -3.03$\pm$0.26 & -9.69$\pm$0.22 & 6485 & 4084 & 10389 & 2109\\
PAR-26 & 4041768185082518784 & - & - & - & - & - & - & -\\
PAR-27$^*$ & 4063251615722434176 & -0.26$\pm$0.15 & -2.23$\pm$0.27 & -3.27$\pm$0.22 & 8313 & 5533 & 12463 & 2097\\
PAR-28 & 4050127570081324672 & -0.09$\pm$0.09 & -2.63$\pm$0.16 & -6.10$\pm$0.15 & 9538 & 6570 & 13998 & 2319\\
PAR-29 & 4050839057193600256 & 0.49$\pm$0.07 & -4.15$\pm$0.14 & -8.88$\pm$0.12 & 1976 & 1711 & 2337 & 2206\\
PAR-30$^*$ & 4062266895673680640 & 0.52$\pm$0.26 & -1.64$\pm$0.46 & -2.22$\pm$0.41 & 2573 & 1386 & 6430 & 2350\\
PAR-31 & 4043716485397203072 & - & - & - & - & - & - & -\\
PAR-32 & 4056556788147423104 & -0.05$\pm$0.20 & -0.91$\pm$0.33 & 2.39$\pm$0.27 & 6356 & 3765 & 10683 & 2346\\
PAR-33$^*$ & 4062443229854120320 & -0.01$\pm$0.17 & -5.33$\pm$0.29 & -6.08$\pm$0.22 & 6309 & 3853 & 10421 & 2262\\
PAR-34$^*$ & 4056339462805179520 & 0.11$\pm$0.27 & -4.21$\pm$0.49 & -4.31$\pm$0.39 & 5090 & 2710 & 9377 & 2376\\
PAR-35 & 4116588783026926848 & 0.78$\pm$0.55 & -1.29$\pm$0.99 & -9.81$\pm$0.80 & 3192 & 1248 & 7111 & 2156\\
PAR-36 & 4050794763198528768 & 0.31$\pm$0.11 & -2.60$\pm$0.18 & -3.82$\pm$0.15 & 3199 & 2224 & 5317 & 2260\\
PAR-38$^*$ & 4043498575920772480 & -0.71$\pm$0.83 & 2.72$\pm$2.08 & -2.32$\pm$1.62 & 5228 & 2560 & 9599 & 2332\\
PAR-39$^*$ & 4056077199264203648 & 0.08$\pm$0.16 & 0.18$\pm$0.28 & -0.58$\pm$0.22 & 5735 & 3417 & 9915 & 2384\\
PAR-42 & 4042375905646767488 & -0.98$\pm$0.26 & -0.29$\pm$0.42 & -6.12$\pm$0.34 & 8844 & 5671 & 13549 & 2336\\
PAR-43 & 4044190413438643584 & - & - & - & - & - & - & -\\
PAR-44 & 4055960822828491008 & -1.35$\pm$0.41 & -2.17$\pm$0.51 & -8.83$\pm$0.37 & 7477 & 4466 & 12100 & 2375\\
PAR-47 & 4055960822828491008 & -1.35$\pm$0.41 & -2.17$\pm$0.51 & -8.83$\pm$0.37 & 7477 & 4466 & 12100 & 2375\\
PAR-48$^*$ & 4050131658891761024 & -1.21$\pm$0.56 & -4.79$\pm$1.24 & -5.73$\pm$1.36 & 6252 & 3471 & 10663 & 2319\\
PAR-50 & 4062953132647842304 & -0.66$\pm$0.24 & -2.41$\pm$0.34 & -3.07$\pm$0.30 & 8009 & 5125 & 12328 & 2170\\
PAR-51 & 4064868889963362048 & -2.15$\pm$0.38 & 3.46$\pm$0.61 & 1.13$\pm$0.52 & 7894 & 5114 & 11962 & 1985\\
PAR-54 & 4062319260040006784 & - & - & - & - & - & - & -\\
PAR-55 & 4053846904343760256 & 0.75$\pm$0.41 & -10.96$\pm$0.65 & -5.07$\pm$0.54 & 2397 & 1084 & 6424 & 2194\\
PAR-56 & 4043455385693259008 & 6.15$\pm$1.17 & -23.34$\pm$2.07 & -4.07$\pm$1.64 & 175 & 138 & 238 & 2342\\
PAR-59 & 4057221236769355136 & -0.12$\pm$0.64 & 1.50$\pm$2.19 & 0.96$\pm$1.75 & 5036 & 2416 & 9412 & 2353\\
\hline
\hline
\end{tabular}
\\$^*$ dark lens candidate 
\end{table*}

\begin{table} 
\begin{scriptsize}
\centering
\caption{OGLE-III parallax events and an estimate of the blending contribution. Events with the blending parameter greater than 0.6 and blend magnitude fainter than 18.5 mag are assumed to have negligible contribution to the light in the Gaia DR2 solution.}
\label{tab:blending}
\begin{tabular}{llllll}
\hline
Lens name  &  solution & $I_0$ & $\fs$ & $I_b$ &  \\
           &  $\u0$/$\piE$ & [mag] &  & [mag] & \\
\hline
\hline 
PAR-01 & -1 +1 & 14.23 & 0.68 & 15.47 & too bright \\
PAR-02 & +1 +1 & 15.45 & 0.64 & 16.56 & too bright \\
PAR-02$^*$ & -1 -1 & 15.45 & 1.01 & - & ok \\
PAR-03$^*$ & +1 +1 & 15.50 & 0.94 & 18.55 & ok \\
PAR-03 & -1 -1 & 15.50 & 0.91 & 18.11 & too bright \\
PAR-04$^*$ & +1 +1 & 14.54 & 1.04 & - & ok \\
PAR-04$^*$ & -1 -1 & 14.54 & 1.03 & - & ok \\
PAR-05$^*$ & +1 +1 & 14.83 & 1.00 & - & ok \\
PAR-05 & -1 -1 & 14.83 & 0.94 & 17.88 & too bright \\
PAR-06 & +1 +1 & 14.72 & 0.46 & 15.39 & too bright \\
PAR-06 & -1 -1 & 14.72 & 0.51 & 15.50 & too bright \\
PAR-07$^*$ & +1 -1 & 15.76 & 0.92 & 18.50 & ok \\
PAR-07 & -1 +1 & 15.76 & 0.71 & 17.10 & too bright \\
PAR-08 & +1 +1 & 15.71 & 0.91 & 18.33 & too bright \\
PAR-08 & -1 -1 & 15.71 & 0.86 & 17.85 & too bright \\
PAR-09$^*$ & +1 -1 & 16.83 & 1.00 & - & ok \\
PAR-11 & +1 +1 & 16.85 & 0.43 & 17.46 & too bright \\
PAR-11 & -1 -1 & 16.85 & 0.52 & 17.64 & too bright \\
PAR-12$^*$ & +1 +1 & 17.44 & 0.66 & 18.61 & ok \\
PAR-12 & -1 -1 & 17.44 & 0.54 & 18.29 & too bright \\
PAR-13$^*$ & +1 -1 & 17.53 & 0.72 & 18.91 & ok \\
PAR-13$^*$ & -1 +1 & 17.53 & 0.73 & 18.95 & ok \\
PAR-14 & +1 +1 & 16.19 & 0.42 & 16.78 & too bright \\
PAR-14 & -1 -1 & 16.19 & 0.17 & 16.40 & too bright \\
PAR-15$^*$ & +1 -1 & 16.83 & 0.87 & 19.05 & ok \\
PAR-15$^*$ & -1 +1 & 16.83 & 0.80 & 18.58 & ok \\
PAR-16 & +1 -1 & 15.80 & 0.79 & 17.49 & too bright \\
PAR-16 & -1 +1 & 15.80 & 1.00 & - & ok \\
PAR-17 & +1 +1 & 17.51 & 0.28 & 17.87 & too bright \\
PAR-17 & +1 -1 & 17.51 & 0.12 & 17.65 & too bright \\
PAR-17 & -1 +1 & 17.51 & 0.20 & 17.75 & too bright \\
PAR-17 & -1 -1 & 17.51 & 0.23 & 17.79 & too bright \\
PAR-19 & +1 -1 & 15.76 & 1.02 & - & ok \\
PAR-19 & -1 -1 & 15.76 & 0.93 & 18.65 & ok \\
PAR-21 & +1 +1 & 16.27 & 0.14 & 16.43 & too bright \\
PAR-21 & -1 -1 & 16.27 & 0.22 & 16.54 & too bright \\
PAR-22$^*$ & +1 +1 & 16.63 & 0.91 & 19.24 & ok \\
PAR-23 & +1 -1 & 16.41 & 0.75 & 17.91 & too bright \\
PAR-23 & -1 +1 & 16.41 & 0.71 & 17.75 & too bright \\
PAR-24$^*$ & +1 +1 & 14.78 & 1.05 & - & ok \\
PAR-24$$ & +1 -1 & 14.78 & 0.62 & 15.83 & too bright \\
PAR-24$$ & -1 +1 & 14.78 & 0.48 & 15.49 & too bright \\
PAR-24 & -1 -1 & 14.78 & 0.88 & 17.09 & too bright \\
PAR-27$^*$ & +1 +1 & 16.49 & 1.08 & - & ok \\
PAR-27 & +1 -1 & 16.49 & 0.57 & 17.40 & too bright \\
PAR-27 & -1 +1 & 16.49 & 0.41 & 17.06 & too bright \\
PAR-27$^*$ & -1 -1 & 16.49 & 0.90 & 18.99 & ok \\
PAR-28 & +1 +1 & 15.24 & 0.96 & 18.74 & ok \\
PAR-28 & -1 -1 & 15.24 & 1.08 & - & ok \\
PAR-29 & +1 -1 & 15.46 & 0.38 & 15.98 & too bright \\
PAR-29 & -1 +1 & 15.46 & 0.23 & 15.74 & too bright \\
PAR-30$^*$ & +1 +1 & 17.26 & 0.71 & 18.60 & ok \\
PAR-30$^*$ & -1 -1 & 17.26 & 0.71 & 18.60 & ok \\
PAR-32 & +1 +1 & 16.54 & 0.16 & 16.72 & too bright \\
PAR-32 & -1 -1 & 16.54 & 0.09 & 16.64 & too bright \\
PAR-33$^*$ & +1 +1 & 16.22 & 1.06 & - & ok \\
PAR-33 & +1 -1 & 16.22 & 0.47 & 16.91 & too bright \\
PAR-33 & -1 +1 & 16.22 & 0.72 & 17.60 & too bright \\
PAR-33 & -1 -1 & 16.22 & 0.86 & 18.36 & too bright \\
PAR-34$^*$ & +1 +1 & 16.76 & 0.93 & 19.65 & ok \\
PAR-34$^*$ & -1 +1 & 16.76 & 0.97 & 20.57 & ok \\
PAR-35 & +1 +1 & 17.62 & 0.56 & 18.52 & too bright \\
PAR-35 & +1 -1 & 17.63 & 0.18 & 17.84 & too bright \\
PAR-35 & -1 +1 & 17.63 & 0.22 & 17.90 & too bright \\
PAR-35 & -1 -1 & 17.62 & 0.53 & 18.44 & too bright \\
PAR-36 & +1 -1 & 15.15 & 0.64 & 16.26 & too bright \\
PAR-36 & -1 -1 & 15.15 & 0.63 & 16.23 & too bright \\
PAR-38$^*$ & +1 +1 & 18.32 & 1.08 & - & ok \\
PAR-38$^*$ & -1 +1 & 18.32 & 1.03 & - & ok \\
\hline
\hline
\end{tabular}
\end{scriptsize}
\\$^*$ dark lens candidate 
\end{table}

\begin{table} 
\begin{scriptsize}
\centering
\caption{Table \ref{tab:blending} continued.  
}
\label{tab:blending2}
\begin{tabular}{llllll}
\hline
Lens name  &  solution & $I_0$ & $\fs$ & $I_b$ &  \\
           &  $\u0$/$\piE$ & [mag] &  & [mag] & \\
\hline
\hline 
PAR-39$^*$ & +1 -1 & 16.32 & 0.90 & 18.82 & ok \\
PAR-39$^*$ & -1 -1 & 16.32 & 0.90 & 18.82 & ok \\
PAR-42 & +1 -1 & 15.94 & 0.92 & 18.68 & ok \\
PAR-42 & -1 +1 & 15.94 & 0.88 & 18.24 & too bright \\
PAR-44 & +1 -1 & 16.88 & 0.84 & 18.87 & ok \\
PAR-44 & -1 +1 & 16.88 & 0.90 & 19.38 & ok \\
PAR-47 & +1 -1 & 16.87 & 0.90 & 19.37 & ok \\
PAR-48$^*$ & +1 +1 & 18.19 & 0.78 & 19.84 & ok \\
PAR-48$^*$ & -1 -1 & 18.19 & 0.69 & 19.46 & ok \\
PAR-50 & +1 +1 & 16.84 & 0.67 & 18.05 & too bright \\
PAR-50 & -1 -1 & 16.84 & 0.34 & 17.30 & too bright \\
PAR-51 & +1 +1 & 15.78 & 0.33 & 16.22 & too bright \\
PAR-51 & -1 +1 & 15.78 & 0.33 & 16.22 & too bright \\
PAR-55 & +1 -1 & 17.51 & 0.21 & 17.77 & too bright \\
PAR-55 & -1 -1 & 17.51 & 0.18 & 17.73 & too bright \\
PAR-56 & +1 +1 & 18.29 & 0.19 & 18.52 & too bright \\
PAR-56 & +1 -1 & 18.29 & 0.16 & 18.47 & too bright \\
PAR-56 & -1 +1 & 18.29 & 0.16 & 18.48 & too bright \\
PAR-56 & -1 -1 & 18.29 & 0.16 & 18.47 & too bright \\
PAR-59 & +1 +1 & 18.17 & 0.28 & 18.52 & too bright \\
PAR-59 & +1 +1 & 18.17 & 0.09 & 18.27 & too bright \\
PAR-59 & +1 -1 & 18.17 & 0.12 & 18.31 & too bright \\
PAR-59 & -1 +1 & 18.17 & 0.15 & 18.34 & too bright \\
PAR-59 & -1 -1 & 18.17 & 0.12 & 18.31 & too bright \\
\hline
\hline
\end{tabular}
\end{scriptsize}
\\$^*$ dark lens candidate 
\end{table}

\section{Dark lens selection}
\citep{Wyrz16} (Wyrz16) analysed light curves of 150 million stars in the Galactic bulge monitored over 8 years by the OGLE-III survey and found 59 long-lasting microlensing events with a significant effect on the amplification due to the Earth orbital motion.   They modelled the light curves with the microlensing parallax model \citep{Smith2005}, which consisted of seven parameters: $\t0$ (time of the maximum amplification), $\te$ (Einstein ring crossing time), $\u0$ (impact parameter), $I_0$ (baseline magnitude in $I$ band), $\fs$ (blending parameter), and $\piEvec$ (microlensing parallax vector with E and N components).


First, we matched all OGLE-III parallax events with the Gaia Data Release 2 catalogue. 
There were 46 matches out of 59 events, with most of the missed ones being fainter than 19 mag in OGLE's $I$ band, hence indicating incompleteness of GDR2 in the bulge region at the faint end of source magnitude distribution. 
Table \ref{tab:gaia} lists all 46 matched events following their names from Wyrz16, along with the GDR2 source identifier (sourceId), Gaia measured parallax, and proper motions in right ascension ($\mu_\mathrm{RA}$) and declination ($\mu_\mathrm{Dec}$).
Five events were identified in GDR2, however, they did not have a full five-parameter astrometric solution provided. 
Again, these were mostly events at the faint end with magnitudes in Gaia around or fainter than 19 mag in $G$ band.

Table \ref{tab:gaia} also lists source distance estimates  $r_\mathrm{est}$ derived from Gaia parallaxes based on the Milky Way priors in 
\cite{Bailer-Jones2018} (hereafter CBJ18) with their lower and upper boundaries (r$_\mathrm{lo}$ and r$_\mathrm{hi}$, respectively, and 
r$_\mathrm{len}$) which is the length of the Galactic prior from CBJ18 used here to derive the posterior distribution of the source distance. 
The posterior distribution on the source distance was used in the later stages of this work to compute the posterior distribution on the lens mass and distance. 
We expect that our sources may be located both in the Galactic disk and the bulge. We use the approach of CBJ18 in distance estimates, although the bulge itself was not properly included in the Milky Way prior in that work. However, as we show in the discussion, any inaccuracy in the distance of the source has a very limited impact on the lens mass estimate. 
We also employ Gaia measurements for sources with estimated negative parallaxes, since the parallax is just a noisy measurement of the inverse distance. In these cases, however, as noted in CBJ18, their distance estimates are strongly affected by the distance distribution prior and consequently have broad posterior distributions. 
Moreover, in most cases with significant negative parallaxes from Gaia, the microlensing models of these events clearly indicate significant blending coming from the lens or a third light source, which affects the Gaia astrometric solution. We filter out these events in the next step.


In order to use GDR2 data on microlensing events and confidently associate the measured astrometric parameters solely with the lensed source, we first had to ensure there is no other significant contribution to the luminosity, whether from a spurious nearby source or from the lens. 
The microlensing parallax model used in Wyrz16 contained a blending parameter $\fs$, which is the fraction of the light contributed to the total baseline of the event (magnitude $I_0$) by the source. 
In the first step, from the sample of 59 parallax events we selected those where the contribution from the blended light was negligible. After this selection we further narrowed down the sample to dark lens events, so this first step is taken primarily to remove obvious strongly blended events. 
We selected those events in which the blended light (non-lensed) in OGLE's $I$ band was fainter than $I_b=I_0-2.5\log_{10}(1-\fs)>18.5$ mag, with the additional requirement that at least 60 \%\ of the light must have originated from the source, $\fs \geq 0.6$.

Events with `negative blending' with $\fs>1$ (e.g., \citealt{Smith2007}) were naturally also included in the sample of sufficiently dark lenses. The constraints on blending in these cases are consistent with the source contributing 100\% of the light and the negative blending, typically at a level of no more than a few percent, can be attributed to the low-level systematics in the data. 


Many of the parallax microlensing events from Wyrz16 had multiple solutions due to known degeneracies \citep{Gould2004, Smith2005}. 
The impact parameter $\u0$ defines the closest approach of the trajectory of the source to the line of sight to the lens. 
In standard non-parallax microlensing events, the relative source-lens motion is closely approximated  by a straight line, hence the problem is symmetric in $\u0$.
In parallax events, the trajectory is sinusoidal and the symmetry is broken; however, the direction of the parallax vector $\piEvec$ can compensate the sign of the impact parameter $\u0$. 
Therefore, a very similar  light curve can be obtained with $\u0$ on either side of the lens, \ie it can be either positive or negative, with different $\piEvec$.
Another degeneracy is related to the jerk motion of the Sun with respect to the Earth, which generates two equally significant solutions for the microlensing parallax vector $\piEvec$. However, as noted in \cite{Gould2004}, this degeneracy depends on the peak moment of the event (with respect to the Earth's position on its orbit) and is most significant for events on timescales shorter than 365/(2$\pi$) days. Therefore, in the Wyrz16 sample of long events only a few events are subject to this degeneracy (\eg PAR-17). In Tables \ref{tab:blending} and \ref{tab:blending2} we denote solutions by indicating the signs of the pairs $\u0$ and $\piEE$ (the eastern component of $\piEvec$)  (\eg +1+1 for $\u0>0$ and $\piEE>0$).

Since each solution could have a different set of other microlensing parameters, in particular the blending parameter $\fs$, we tested each solution separately. In many cases all solutions were qualified as sufficiently dark; however, there were also individual solutions which we had to exclude (\eg +1-1 and -1+1 solutions for PAR-27). Table \ref{tab:blending} lists the blending parameter and magnitude of the blend for each of the solutions for all events, along with the decision ({\it ok} or {\it too bright}).
For example, the $\u0<0$ solution for PAR-02, the most massive of the dark lenses from Wyrz16, returned a negligible blending contribution, while the $\u0>0$ solution had a very bright blend of $I_b=16.5$ mag for $\fs=0.64$. If the latter solution was valid, the presence of a blend could have affected the Gaia measurement of parallax and proper motions, providing an unreliable astrometric solution \citep{Bramich2018}.   On the other hand, as shown in Wyrz16, for the blended solution for PAR-02 the expected amount of blending light from the lens if the lens were a main sequence star is still significantly greater than the observed blended light; therefore, the blended solution still yields a black hole candidate. 

Therefore, for further analysis we only used microlensing events with solutions that indicated a low contribution of the blended light.  For each individual solution below we compute the dark-lens probability by comparing the blended light to the expected  main sequence star luminosity for the computed lens mass and we select the one with the highest probability.  The subsequent tables in this work only display the solutions selected at this stage. The remaining solutions are available in Wyrz16.  In total, out of 59 events there remain 24 with GDR2 data and at least one solution with a negligible blending contribution. 


\subsection{Dark lens probability}
In order to improve the selection on dark lenses we repeat the procedure of Wyrz16, as summarised below, and use their full Markov chain Monte Carlo (MCMC) samples for microlensing parallax models. 

For a standard microlensing event it is typically impossible to obtain a unique solution for the distance and mass of the lens, since there is only one physical parameter ($\te$) obtained from the observed light curve, which is determined by a combination of the source and lens distance, the lens mass and the lens--source relative velocity. 
However, for parallax events the situation improves as we additionally measure the microlensing parallax $\piE=(\piL-\piS)/\thetaE$, where $\piL=1/\DL$ and $\piS=1/\DS$ are parallaxes (distances) of the lens and the source, respectively.  The mass and distance of the lensing object can then be derived as

\begin{equation}
M=\frac{\theta_\mathrm{E}}{\kappa \pi_\mathrm{E}} = \frac{\mu_\mathrm{rel} \tE}{\kappa \pi_\mathrm{E}}
\end{equation}
and
\begin{equation}
D_{\rm{L}}=\frac{1}{\mu_{\rm{rel}} \tE \pi_{\rm{E}} + 1/D_{\rm{S}}}\, ,
\end{equation}where we used the fact that the angular size of the Einstein radius can be rewritten as a product of the length of the vector of the heliocentric relative proper motion $|\mu_\mathrm{rel}|=|\mu_L-\mu_S|$ 
between lens (L) and source (S) and the event's timescale $\te$. 
Proper motions and the timescale should be measured in the same frame, either geocentric or heliocentric. We therefore converted the timescales of events obtained in our geocentric parallax model into the heliocentric frame, following \cite{Skowron2011}. 
In other words, the motion of the Earth was taken out of the relative proper motion computation, hence if the lens was observed from the Sun, it would have the timescale of $\tEhelio$ (as obtained in the models in Wyrz16), hereafter denoted $\tE$.

For each sample we computed posterior probability distributions on the mass and distance of the lens using the probability distribution on the distance to the source from CBJ18 and source proper motion $\muS$ and its error estimates from GDR2.  We sample the lens proper motion $\muL$ from the disk stars distribution following \cite{Calamida2014}, 
which assumes that black holes do not receive additional natal kicks (see section 5).
We also exploit the fact that the direction of the $\muL$ vector is aligned within the error bar with the vector of $\piEvec=(\piEE, \piEN)$. 

For each MCMC sample from which we derived the mass and distance we also computed the weight (equivalent of a prior) on the parameters, following Wyrz16 and \cite{Batista2011}. We used here a prior on lens and source Galacto-centric three-dimensional positions as well as proper motions following the distributions for the bulge and disk for the source and the lens, respectively. 

Once the mass of the lens and its distance are computed for each MCMC sample, we computed its expected extinction-free observed brightness if it was a regular main sequence star,  using the following luminosity-mass relations:
\[
I_\mathrm{lens} =
\begin{cases}
 14.9 \mathrm{~mag} - 2.5 \log{\frac{\ML^4}{\DL^2}} & \quad \text{for } \ML<2 \msun \\
 14.9 \mathrm{~mag} - 2.5 \log{\frac{1.5 \ML^{3.5}}{\DL^2}} & \quad \text{for } \ML>2 \msun
\end{cases}
\]

Then, for the same sample using $\Izero$ and $\fs$ and the extinction derived for OGLE-III fields from \cite{Nataf}, we computed the amount of light which could be coming from the blend if all the light were attributed to the lens. This yields an upper limit on the luminosity of the lens since the blending light could be split between the lens and an unrelated light source within the seeing disk of the OGLE image. 

The overall probability that a lens in a given event is dark is computed as the fraction of MCMC samples for which the blending luminosity was insufficient to accommodate the luminosity inferred from the mass and distance of the lens. Table \ref{tab:probs} lists the probabilities obtained for all 24 events and their individual solutions. By selecting solutions with dark lens probability higher  than 75 \%\ (as in Wyrz16) we construct our sample of 18 events with likely dark lenses. These events are marked with an asterisk ($*$) in all the tables in this work. 
We note that some events had two or more degenerate solutions, and probabilities were derived for each of them separately as they yielded different blending parameters, masses, and distances. Therefore, for our further analysis of the mass distribution we used only the most probable dark solution for each of the events.  

We estimated the mass and distance of the lens from the resulting posterior probability by marginalising over the remaining parameters.
Table \ref{tab:massesdistances} lists median estimates for the mass and distance of the lens and the brightness of the blend for all events with Gaia DR2 astrometric parameters. 
The large  blend magnitudes shown, such as $\sim$21.95 mag, are not actual measurements, but indicate that the blending parameter $\fs$ was close to or exceeded 1.0 (negative blending), meaning there is no additional light present. 

\subsection{Colour-magnitude diagram}
In Wyrz16 the distance to the source was not known, hence for dark lens candidates we only selected among the events with sources located within the Red Clump region and assumed their distance as 8 kpc. 
In this work we can relax this constraint and include sources outside of the Red Clump region when GDR2/CBJ18 data are available. 

\begin{figure} 
\centering
\includegraphics[width=8.5cm]{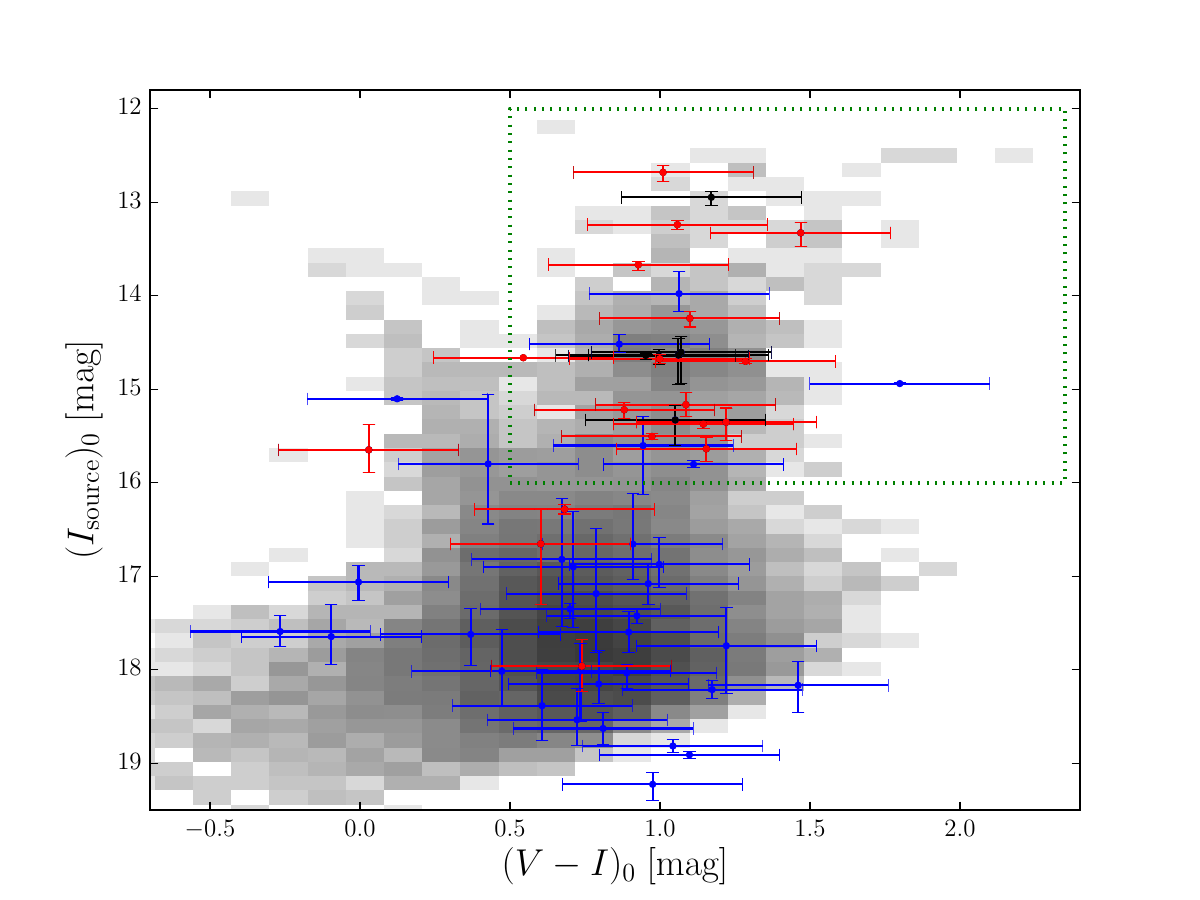}
\caption{Extinction-free colour-magnitude diagram for sources with parallax events from Wyrz16. The background is the extinction-corrected CMD of stars in the  sub-field of the PAR-02 event.
Selected dark lenses are marked in red. 
Black indicates sources in events with GDR2 data selected after the blending cut where the lens is consistent with being a main sequence star.
Blue indicates the remaining events. 
The green dotted box indicates the Red Clump Region used in Wyrz16.} 
\label{fig:cmd}
\end{figure}

Figure \ref{fig:cmd} shows a colour-magnitude diagram (CMD) with all 59 parallax events from Wyrz16 with the 18 selected dark lenses marked in red. Black points mark the remaining six events for which there were GDR2 data and which passed the blending criteria; however, their dark-lens probability was below 75\%. The remaining events (too bright blend) are marked in blue. 
The $I$-band magnitudes of the sources were de-blended using $\Izero$ and $\fs$ from the microlensing parallax MCMC models and were corrected for extinction, based on \cite{Nataf} extinction maps for OGLE-III. 
The background shows the stars from the field containing PAR-02 event, as a typical Bulge field, for reference only. 
There are two events (PAR-12 and PAR-30) with sources located outside of the Red Clump region for which the use of Gaia distances and proper motion indicated the dark nature of the lens. 


\begin{figure}
\centering
\includegraphics[width=11.5cm]{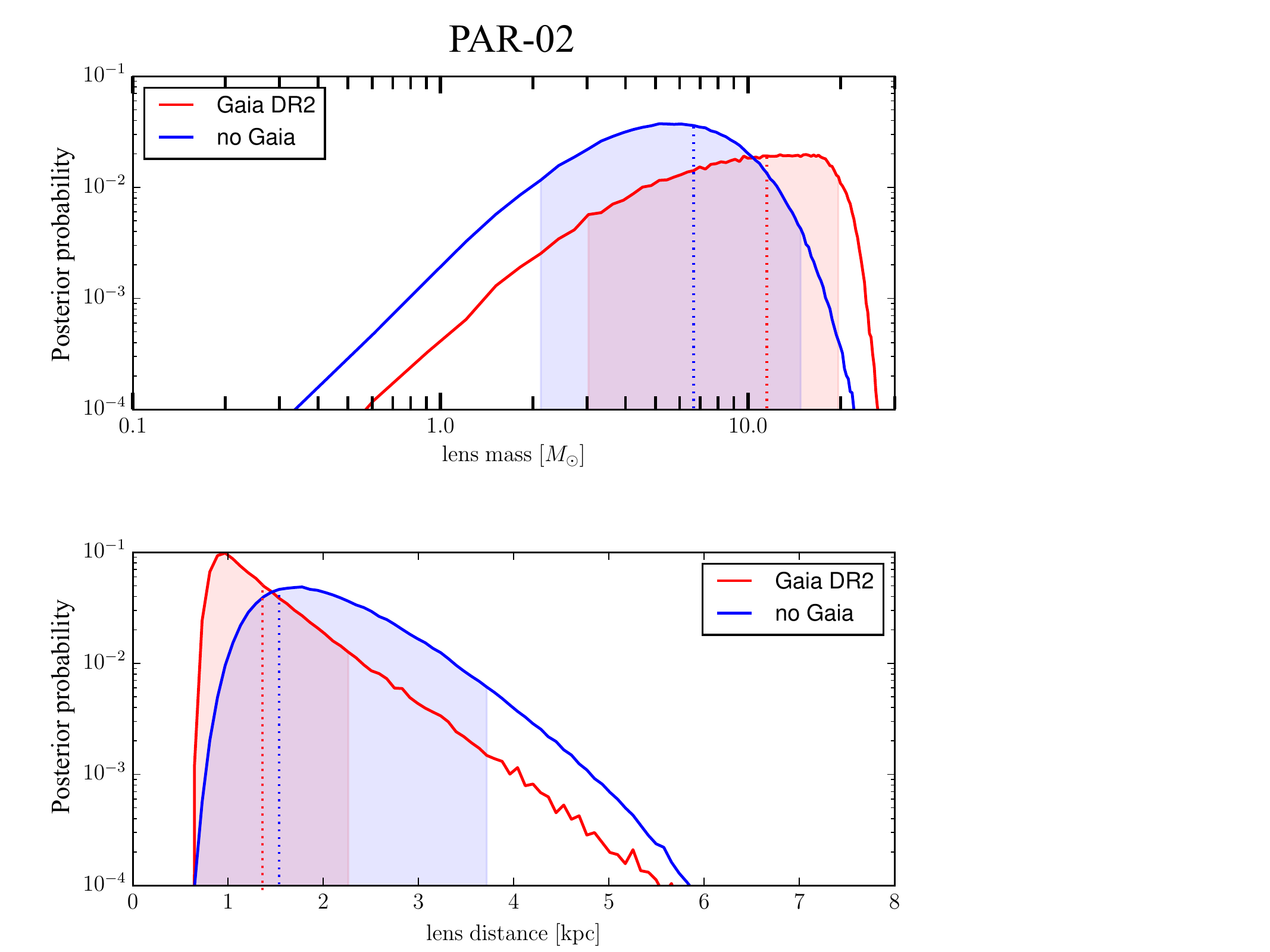}
\caption{
Comparison between mass (top) and distance (bottom) posterior probability densities derived with (red) and without (blue) Gaia DR2 information on the source for the most massive dark lens in our sample, OGLE3-ULENS-PAR-02. Dotted lines indicate medians, and shaded regions denote the 95\% confidence level. 
} 
\label{fig:gaia-nogaia}
\end{figure}

\begin{table} 
\begin{scriptsize}
\centering
\caption{Probabilities for having a dark lens in parallax events. 
Probabilities are shown for all solutions in which the blend is fainter than 18.5 mag and events are sorted according to the highest probability of any of the solutions. PAR-34 is the last candidate included above the 75 \%\ threshold. Dark lens candidates are flagged with asterisks.}
\label{tab:probs}
\begin{tabular}{lllll}
\hline
Lens name  &  $u_0>0$ & $u_0>0$ & $u_0<0$ & $u_0<0$ \\
OGLE3-ULENS- &  $\piEN>0$ & $\piEN<0$ & $\piEN>0$ & $\piEN<0$ \\
\hline
\hline 
PAR-02$^*$ & - & - & - & 99.9 \\
PAR-13$^*$ & - & 99.9 & 99.8 & - \\
PAR-05$^*$ & 99.5 & - & - & - \\
PAR-04$^*$ & 99.1 & - & - & 99.2 \\
PAR-07$^*$ & - & 99.2 & - & - \\
PAR-30$^*$ & 99.1 & - & - & 98.8 \\
PAR-15$^*$ & - & 98.3 & 95.4 & - \\
PAR-12$^*$ & 95.7 & - & - & - \\
PAR-03$^*$ & 94.5 & - & - & - \\
PAR-09$^*$ & - & 94.4 & - & - \\
PAR-33$^*$ & 93.2 & - & - & - \\
PAR-38$^*$ & 92.8 & - & 91.8 & - \\  
PAR-27$^*$ & 90.8 & - & - & 77.7 \\
PAR-39$^*$ & - & 88.8 & - & 88.5 \\
PAR-48$^*$ & 88.3 & - & - & 85.7 \\
PAR-22$^*$ & 83.1 & - & - & - \\
PAR-24$^*$ & 77.9 & - & - & - \\ 
PAR-34$^*$ & 76.9 & - & 80.2 & - \\ 
\hline
PAR-19 & - & 70.8 & - & 69.2 \\
PAR-28 & 55.4 & - & - & 69.6 \\
PAR-44 & - & 55.1 & 48.8 & - \\
PAR-42 & - & 45.8 & - & - \\
PAR-16 & - & - & 41.2 & - \\
PAR-47 & - & 30.4 & - & - \\
\hline
\hline
\end{tabular}
\end{scriptsize}
\end{table}

\begin{table} 
\begin{small}
\centering
\caption{Posterior medians and 1$\sigma$ uncertainty estimates on masses, distances, and total blend magnitudes for all events matched with Gaia DR2. Multiple entries are shown for different solutions found, indicated in the last column. Blend magnitudes fainter than 21.9 mag correspond to solutions with negative blending, consistent with no blending. 
}
\label{tab:massesdistances} 
\begin{tabular}{lllll}
\hline
Lens name  &  mass & distance & blend & solution \\
OGLE3-ULENS- &  [$\msun$] & [kpc] & [mag] & ($\u0$ $\piEN$)\\
\hline
\hline 
PAR-02$^*$ & $11.9^{+4.9}_{-5.2}$ & $1.3^{+0.7}_{-0.3}$ & $21.93^{+0.00}_{-4.99}$ & -1 -1 \\
PAR-03$^*$ & $2.4^{+1.9}_{-1.3}$ & $0.6^{+0.5}_{-0.2}$ & $17.69^{+0.53}_{-0.32}$ & +1 +1 \\
PAR-04$^*$ & $2.9^{+1.4}_{-1.3}$ & $0.7^{+0.4}_{-0.2}$ & $21.93^{+0.00}_{-3.85}$ & +1 +1 \\
PAR-04$^*$ & $3.2^{+1.3}_{-1.3}$ & $0.7^{+0.4}_{-0.2}$ & $21.93^{+0.00}_{-4.50}$ & -1 -1 \\
PAR-05$^*$ & $6.7^{+3.2}_{-2.7}$ & $1.6^{+0.7}_{-0.4}$ & $19.53^{+2.39}_{-4.22}$ & +1 +1 \\
PAR-07$^*$ & $4.5^{+1.8}_{-1.8}$ & $1.5^{+0.7}_{-0.3}$ & $17.16^{+0.65}_{-0.41}$ & +1 -1 \\
PAR-09$^*$ & $1.9^{+1.0}_{-1.0}$ & $0.8^{+0.6}_{-0.2}$ & $21.94^{+0.00}_{-4.01}$ & +1 -1 \\
PAR-12$^*$ & $2.7^{+1.2}_{-1.2}$ & $0.9^{+0.5}_{-0.3}$ & $16.33^{+0.89}_{-0.38}$ & +1 +1 \\
PAR-13$^*$ & $9.0^{+3.9}_{-3.7}$ & $1.9^{+0.9}_{-0.5}$ & $16.76^{+0.22}_{-0.22}$ & +1 -1 \\
PAR-13$^*$ & $8.0^{+4.1}_{-3.2}$ & $1.8^{+0.8}_{-0.4}$ & $16.78^{+0.18}_{-0.16}$ & -1 +1 \\
PAR-15$^*$ & $3.6^{+1.4}_{-1.6}$ & $1.2^{+0.6}_{-0.3}$ & $17.11^{+1.69}_{-0.54}$ & +1 -1 \\
PAR-15$^*$ & $2.8^{+1.3}_{-1.3}$ & $1.3^{+0.6}_{-0.4}$ & $16.80^{+0.73}_{-0.42}$ & -1 +1 \\
PAR-16 & $0.4^{+0.5}_{-0.2}$ & $1.2^{+1.5}_{-0.6}$ & $21.94^{+0.00}_{-4.52}$ & -1 +1 \\
PAR-19 & $2.0^{+1.6}_{-0.9}$ & $4.9^{+1.9}_{-1.7}$ & $21.93^{+0.00}_{-5.86}$ & +1 -1 \\
PAR-19 & $2.1^{+1.4}_{-1.1}$ & $2.8^{+1.8}_{-1.1}$ & $15.65^{+0.90}_{-0.45}$ & -1 -1 \\
PAR-22$^*$ & $1.2^{+0.7}_{-0.5}$ & $1.0^{+0.7}_{-0.3}$ & $17.61^{+4.33}_{-0.96}$ & +1 +1 \\
PAR-24$^*$ & $1.4^{+1.3}_{-0.7}$ & $2.8^{+1.9}_{-1.1}$ & $17.71^{+4.22}_{-1.88}$ & +1 +1 \\
PAR-27$^*$ & $0.9^{+0.8}_{-0.5}$ & $1.8^{+1.3}_{-0.7}$ & $21.95^{+0.00}_{-0.00}$ & +1 +1 \\
PAR-27$^*$ & $1.1^{+0.7}_{-0.6}$ & $1.2^{+1.3}_{-0.5}$ & $17.89^{+0.73}_{-0.42}$ & -1 -1 \\
PAR-28 & $0.9^{+0.6}_{-0.5}$ & $2.9^{+1.5}_{-1.0}$ & $21.94^{+0.00}_{-4.27}$ & +1 +1 \\
PAR-28 & $0.8^{+0.5}_{-0.4}$ & $1.7^{+1.2}_{-0.5}$ & $21.95^{+0.00}_{-0.00}$ & -1 -1 \\
PAR-30$^*$ & $3.0^{+1.3}_{-1.2}$ & $1.3^{+0.6}_{-0.3}$ & $18.00^{+3.94}_{-1.00}$ & +1 +1 \\
PAR-30$^*$ & $2.8^{+1.1}_{-1.0}$ & $1.2^{+0.5}_{-0.3}$ & $17.81^{+4.14}_{-0.88}$ & -1 -1 \\
PAR-33$^*$ & $1.2^{+0.8}_{-0.5}$ & $2.0^{+1.3}_{-0.7}$ & $21.96^{+0.00}_{-0.00}$ & +1 +1 \\
PAR-34$^*$ & $2.0^{+1.2}_{-0.9}$ & $2.5^{+1.1}_{-0.8}$ & $16.41^{+5.51}_{-1.20}$ & +1 +1 \\
PAR-34$^*$ & $2.1^{+1.2}_{-0.9}$ & $2.6^{+1.2}_{-0.9}$ & $16.97^{+4.95}_{-1.52}$ & -1 +1 \\
PAR-38$^*$ & $1.0^{+0.7}_{-0.5}$ & $1.5^{+1.0}_{-0.6}$ & $21.97^{+0.00}_{-0.00}$ & +1 +1 \\
PAR-38$^*$ & $1.1^{+0.8}_{-0.6}$ & $1.7^{+1.1}_{-0.6}$ & $21.97^{+0.00}_{-0.00}$ & -1 +1 \\
PAR-39$^*$ & $2.6^{+1.7}_{-1.3}$ & $3.6^{+2.3}_{-1.5}$ & $17.05^{+0.05}_{-0.05}$ & +1 -1 \\
PAR-39$^*$ & $2.3^{+1.9}_{-1.0}$ & $3.4^{+2.5}_{-1.4}$ & $17.04^{+0.05}_{-0.05}$ & -1 -1 \\
PAR-42 & $0.6^{+0.4}_{-0.3}$ & $1.8^{+1.3}_{-0.6}$ & $18.20^{+3.74}_{-1.31}$ & +1 -1 \\
PAR-44 & $1.2^{+0.8}_{-0.6}$ & $2.7^{+1.6}_{-1.0}$ & $16.48^{+5.45}_{-1.07}$ & +1 -1 \\
PAR-44 & $0.9^{+0.7}_{-0.5}$ & $2.5^{+1.3}_{-0.9}$ & $17.19^{+4.74}_{-1.34}$ & -1 +1 \\
PAR-47 & $0.6^{+0.4}_{-0.3}$ & $1.8^{+1.2}_{-0.6}$ & $17.00^{+4.92}_{-1.36}$ & +1 -1 \\
PAR-48$^*$ & $1.4^{+1.1}_{-0.7}$ & $2.1^{+1.1}_{-0.8}$ & $18.84^{+0.93}_{-0.39}$ & +1 +1 \\
PAR-48$^*$ & $1.3^{+1.1}_{-0.7}$ & $2.1^{+1.2}_{-0.7}$ & $18.68^{+0.68}_{-0.28}$ & -1 -1 \\
\hline
\hline
\end{tabular}
\end{small}
\end{table}

\section{Mass function}

\begin{figure}
\centering
\includegraphics[width=8.5cm]{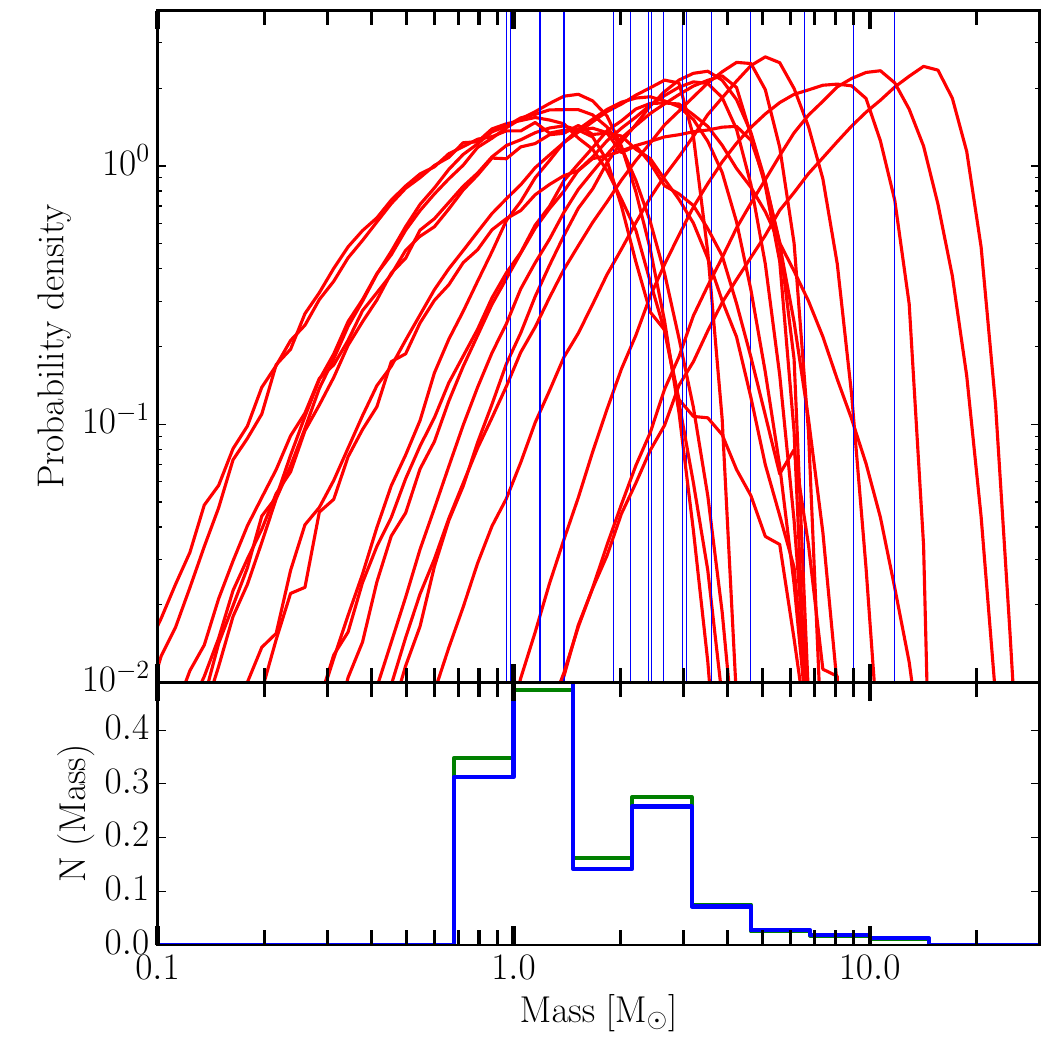}
\caption{Upper panel: Mass posterior probability densities for 18 dark lenses in OGLE parallax events with Gaia DR2 parameters. Blue vertical lines indicate their medians listed in Table \ref{tab:massesdistances}. 
Lower panel: Histogram of median masses in green before efficiency correction and in blue after correction for detection efficiency.}
\label{fig:mass}
\end{figure}

Two-dimensional posterior probability density functions in the mass-distance space were obtained for each event.  These can be marginalised over distance to derive the one-dimensional probability densities for the dark-lens masses, or over mass to generate  the distances.
To illustrate the impact of the Gaia data, we show mass and distance posteriors for the PAR-02 lens obtained with and without Gaia DR2 information (as in Wyrz16) in Fig.~\ref{fig:gaia-nogaia}. 
Figure \ref{fig:mass} shows the mass probability density functions for all 18 events (their `darkest' solution)  computed using GDR2 parameters for the sources (upper panel). 
The thin blue lines show the median values for each mass. The lower panel is the histogram of median masses of this sample with (blue) and without (green) efficiency correction, using the same efficiency correction as in Wyrz16; this panel is shown for illustration only, and the full mass posteriors are used for the population modelling as described below.

We use hierarchical Bayesian inference to infer the mass function of compact objects from the data set described above \citep{Hogg:2010}.  We apply the methodology of \citet{Mandel:2010stat}, who specifically considered
inference on a mass distribution given a sample of uncertain measurements.

Let the predicted mass distribution be described by a model with parameters $\vec{\theta}$ (\ie the model defines $p(M|\vec{\theta})$).  The probability of making a set of observations $\vec{d}$ is 
\begin{equation}
p(\vec{d}|\vec{\theta}) = \prod_i^{N_\mathrm{obs}} p(d_i|\vec{\theta}),
\end{equation}
where $N_\mathrm{obs}$ is the number of independent observations and the probability of making an individual observation is given by
\begin{equation}
p(d_i|\vec{\theta}) = \int dM p(d_i|M) p(M|\vec{\theta}).
\end{equation}
This integral over all possible values of the true mass corresponding to an individual observation can be evaluated with a sum over samples of mass drawn according to the likelihood function $p(d_i|M)$.  

Once $p(\vec{d}|\vec{\theta})$ is obtained, we can use Bayes' theorem to transform it into a posterior on the model parameters $\vec{\theta}$, up to a normalising evidence $p(\vec{d})$: 
\begin{equation}
p(\vec{\theta}|\vec{d}) = \frac{p(\vec{\theta})p(\vec{d}|\vec{\theta})}{p(\vec{d})}.
\end{equation}
We use flat priors on the model parameters $p(\vec{\theta})$.  
This does not account for selection effects \citep{Mandel:2018select}; however, these are roughly constant across the mass space in our sample of microlensing remnants, and do not impact the results below. 

It is possible to consider a set of models and use the ratio of their evidence $p(\vec{d}|\textrm{Model})$ for model selection.  This was the approach taken by \citet{Farr:2010} when analysing the mass distribution of black holes in X-ray binaries.  However, the evidence is sensitive to a number of arbitrary choices, such as the prior range (e.g., a broader prior that extends beyond the region of parameter space which has likelihood support reduces the evidence).  Given the limited data set, significant error bars, and the lack of confident physical models to compare between, we eschew model selection in favour of a simple phenomenological model that allows for the possibility of a mass gap between neutron stars and black holes.

Our population mass model consists of an admixture of a neutron-star population and a black hole population.  The neutron stars (and any high-mass white dwarfs that may be mixed into the sample) are assumed to follow a flat mass distribution between 1 and 2 solar masses.  We have tested that we are not very sensitive to the assumed shape of this distribution.  The black holes are assumed to follow a power-law distribution in mass $dN/dM \propto M^\lambda$, starting with a minimum mass $M_\mathrm{min}$.  If $M_\mathrm{min} = 2 M_\odot$, there is no mass gap; otherwise, there is a gap between $2 M_\odot$ and $M_\mathrm{min}$.  The mixing ratio between the neutron star (NS) and black hole populations is a free parameter $\alpha$.  Thus, the assumed normalised mass distribution is

\begin{eqnarray}
p(M|\alpha,M_\mathrm{min},\lambda) = (\alpha-1) M^\lambda \frac{\lambda+1}{M_\mathrm{min}^{\lambda+1}}&& \mathrm{for}\, M \geq M_\mathrm{min} \geq 2\ M_\odot\,\\
p(M|\alpha,M_\mathrm{min},\lambda) = \frac{\alpha}{M_\odot}&& \mathrm{for}\, 2\ M_\odot \geq M_\mathrm{min} \geq 1\ M_\odot\, \nonumber\\
p(M|\alpha,M_\mathrm{min},\lambda)=0&& \mathrm{otherwise}\, \nonumber.
\end{eqnarray}

There are three free parameters that comprise the model specification $\vec{\theta}$: $\alpha \in [0,1]$, $M_\mathrm{min} \geq 2 M_\odot$, and $\lambda$.  We evaluate the posterior probability density function on a grid over these three parameters. 

\begin{figure}
\centering
\includegraphics[width=8.5cm]{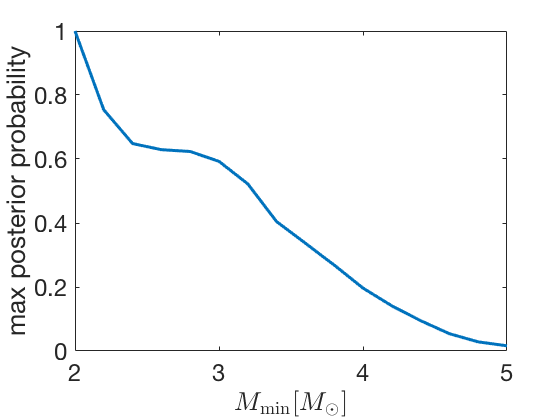}
\caption{Maximum posterior for a given choice of the minimum of the black hole mass distribution $M_\mathrm{min}$, maximised over $\lambda$ and $\alpha$, normalised so that the absolute maximum posterior equals 1.} 
\label{fig:Mmin}
\end{figure}

Figure \ref{fig:Mmin} shows the value of the posterior maximised over $\lambda$ and $\alpha$ as a function of $M_\mathrm{min}.$  The posterior has a maximum at $M_\mathrm{min} = 2 M_\odot$, $\lambda=-3.2$, and $\alpha=0$.   This means that we prefer a single power law slope $p(M) \propto M^{-3.2}$ for $M \geq2 M_\odot$, and 0 elsewhere.  

The overall posterior maximum lies at $\alpha=0$, which suggests no preference for a distinct NS sub-population in the data.  However, if $\alpha$ is fixed to $0.5$, forcing a 50-50 ratio between NSs and BHs, the maximum posterior (which now lies at $M_\mathrm{min}=3 M_\odot$) only decreases by a factor of 2 relative to $\alpha=0$.   
Thus, we find that $\alpha=0$ is not a significant preference and the branching ratio between NS and BH lenses is not well constrained by our data.

Population analyses of black hole masses in X-ray binaries \citep{Ozel2010,Farr:2010} indicate that the black hole mass distribution in black hole X-ray binaries occupies the mass range above $5 M_\odot$.
We looked for a corresponding wide gap between neutron star masses below $2 M_\odot$ and black hole masses starting with $5 M_\odot$ by setting $M_\mathrm{min}=5 \msun$ in our model.  We find that a wide mass gap is strongly disfavoured, with a maximum posterior for $M_\mathrm{min}=5 \msun$ that is a factor of 60 lower than the maximum posterior found for the no-gap model.

\begin{figure}
\centering
\includegraphics[width=8.5cm]{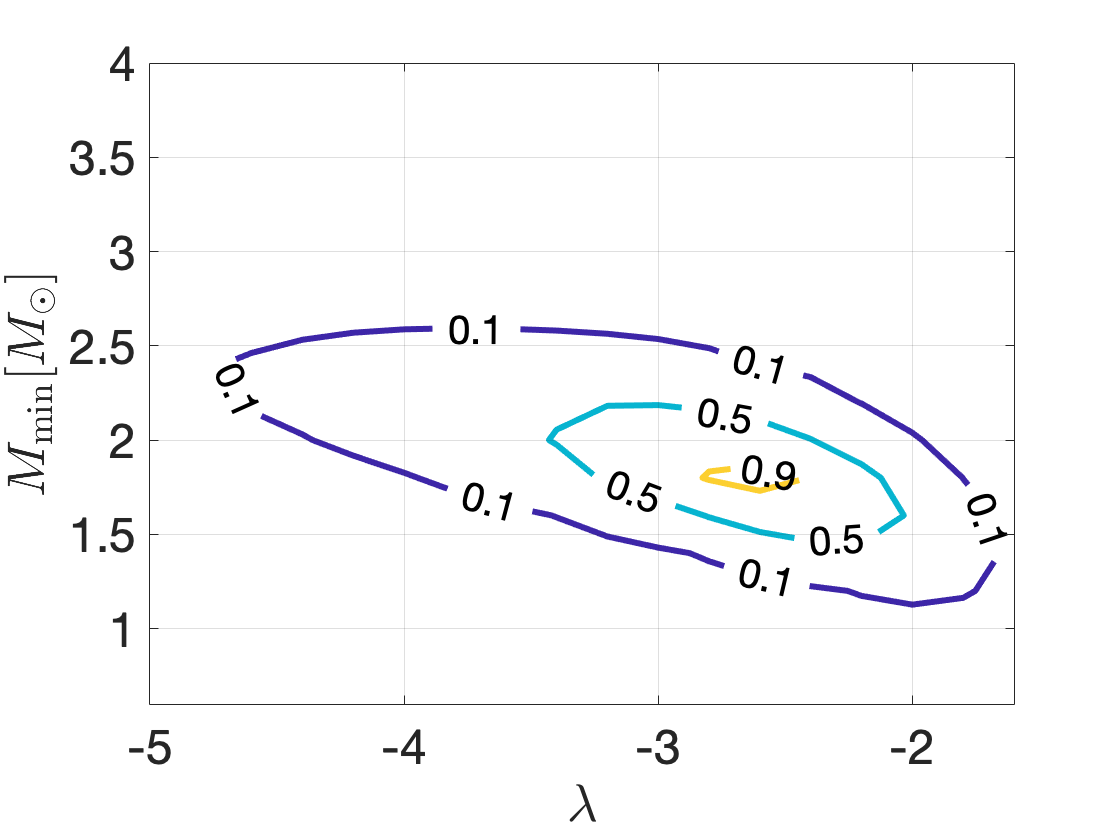}
\caption{Posterior probability as a function of $M_\mathrm{min}$ and the mass distribution power-law index $\lambda$ for a single power-law model, normalised so that the absolute maximum posterior equals 1.} 
\label{fig:lambdaMmin}
\end{figure}

As an alternative model, we consider a single power law $dN/dM \propto M^\lambda$, starting with a minimum  mass $M_\mathrm{min}$, and eschewing a distinct sub-population of neutron stars with a flat mass distribution.  The two-dimensional contour plot in Figure \ref{fig:lambdaMmin} shows the shape of the posterior surface in the $(M_\mathrm{min},\lambda)$ space.   The posterior is peaked at $M_\mathrm{min} = 1.8 M_\odot$ and $\lambda=-2.7$.  High values of $M_\mathrm{min}$ typical for assumed mass gaps are disfavoured in this simpler two-parameter model just as in the three-parameter model considered above.  Meanwhile, the limited support for low $M_\mathrm{min}$ values indicates that the earlier conclusions are robust to the chosen prior boundary on $M_\mathrm{min}$.

We thus conclude that there is no evidence for a mass gap in this sample, and strong evidence against a wide gap spanning the range from $2 M_\odot$ to $5 M_\odot$.

\section{Mass-gap objects}

%


As argued in the previous section, our microlensing data do not support the existence of a mass gap between neutron star and black hole masses.  Here we consider specific events with estimated masses overlapping the putative mass gap range between 2 and 5 $\msun$.
Our revised sample of dark lenses from OGLE-III parallax events contained eight such dark lens events, whose  microlensing parameters were presented in Wyrz16. 

\begin{figure*}
\centering
\includegraphics[width=18cm]{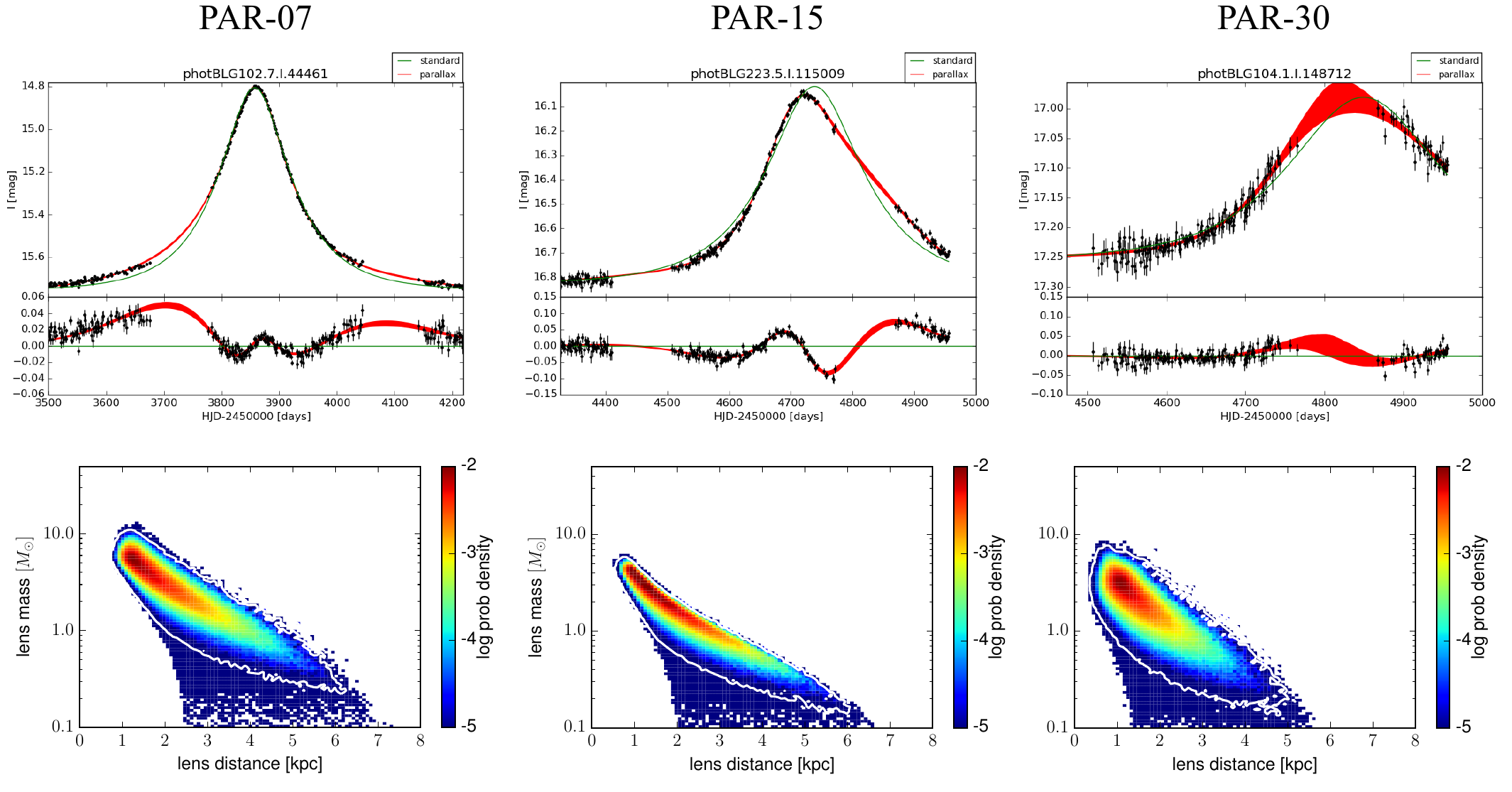}
\caption{Microlensing events with mass estimates in the mass gap between 2 and 5 $\msun$.  Top panel: Light curves with the standard model (green) and annual parallax model (red) derived with MCMC modelling, with residuals relative to the standard model plotted below. Bottom panel: Mass-distance posterior probability densities along with the 95\% confidence level contour.}
\label{fig:mass-gap-objects}
\end{figure*}

Figure \ref{fig:mass-gap-objects} shows the light curves and probability density maps for mass and distance of the lens after including Gaia DR2 information for three of these events.

OGLE3-ULENS-PAR-07 (EWS: OGLE-2005-BLG-474) had two solutions for $\u0$ in Wyrz16 with different blending parameters (0.92 and 0.71, for + and - solutions, respectively). Because of our requirement on the absence of significant light contamination in order to reliably use GDR2 data, the more blended solution was not included here.  The mass of the lens was estimated as $4.5^{+1.8}_{-1.8}$ $\msun$. In Wyrz16 both solutions yielded masses of about 3 $\msun$, with larger error bars.

OGLE3-ULENS-PAR-15 (EWS: OGLE-2008-BLG-096) also had two solutions for $\u0$ in the microlensing parallax model.  Both of them  satisfied the blending condition. 
The mass was estimated as $3.6^{+1.4}_{-1.6}$ $\msun$ and  $2.8^{+1.3}_{-1.3}$ $\msun$, for the + and - solutions, respectively, with the + solution having a slightly higher probability for being dark (98.3\%\ versus 95.4\%, respectively). The previous estimate for the mass of this event in Wyrz16 was between 1.6 and 2.1 $\msun$.We comment on the trend for higher mass estimates than in Wyrz16 below.

OGLE3-ULENS-PAR-30 (EWS: OGLE-2008-BLG-545) also had two solutions, both with a fairly large amount of blending (0.71). However, with a faint baseline of 17.2 mag, the contaminating contribution is again below 18.5 mag. 
This event was not present in the dark lens sample of Wyrz16, as it was located outside of the Red Clump region, hence the distance to the source was not known before Gaia's DR2. 
The masses were computed as $3.0^{+1.3}_{-1.2}$ $\msun$ and $2.8^{+1.1}_{-1.0}$ $\msun$, for the + and - solutions, respectively. 

\begin{figure}
\centering
\includegraphics[width=8.5cm]{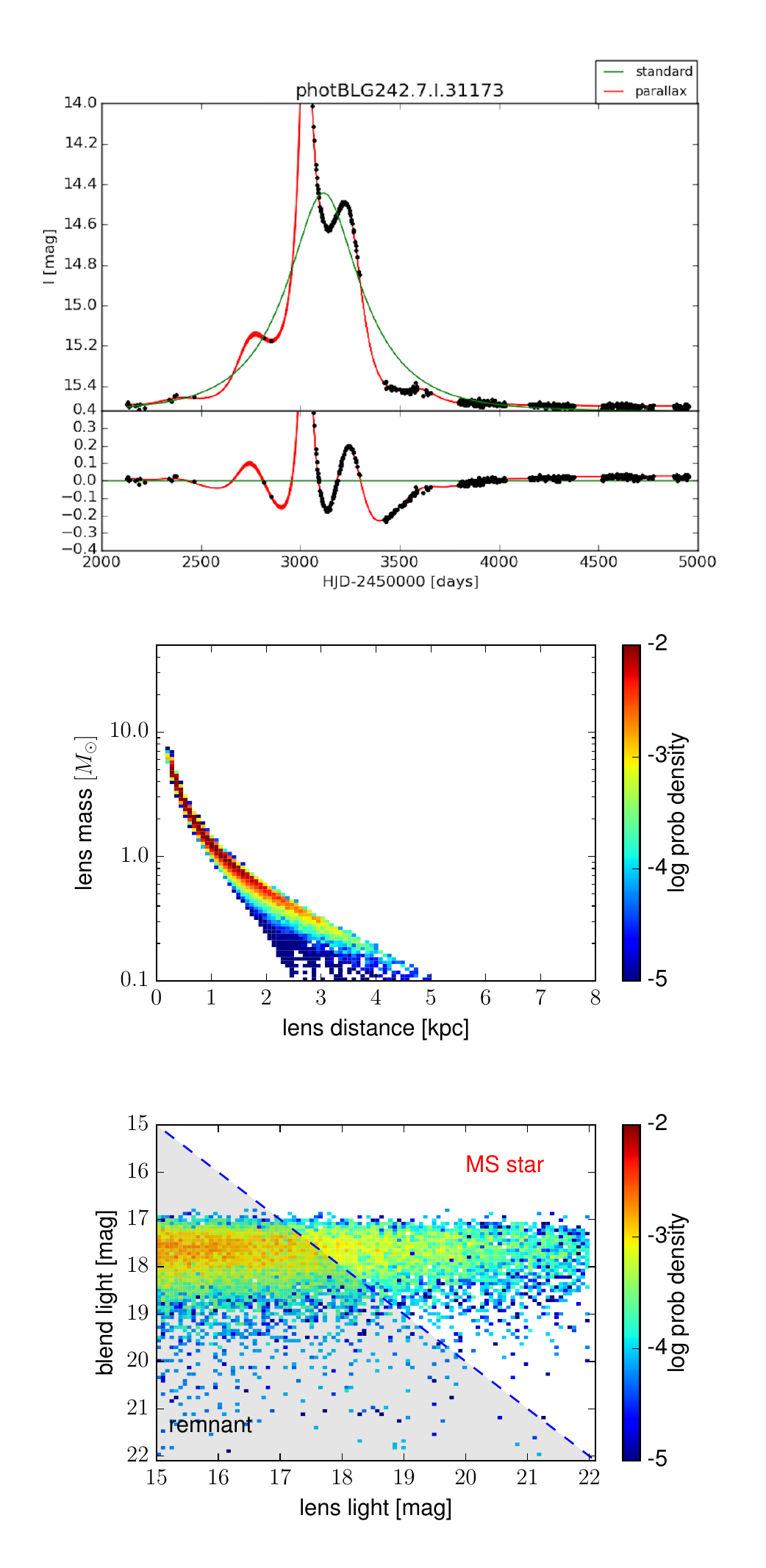}
\caption{Light curve (top) and probability map for mass and distance of the lens (middle) and blended and/or lensed light (bottom) for the event OGLE3-ULENS-PAR-03. This is the nearest ($\sim$ 600 pc) possible mass-gap object in our sample, with a mass of $2.4^{+1.9}_{-1.3}$ $\msun$. The probability that the lens is a dark remnant is 94.5 \%\ based on the integrated histogram from the bottom panel.} 
\label{fig:par-03}
\end{figure}

Another event, OGLE3-ULENS-PAR-03 (not found by OGLE's EWS), has an estimated mass of $2.4^{+1.9}_{-1.3}$ $\msun$, which lies at the edge of the putative mass gap and could still be consistent with a NS or even a White Dwarf lens. 
Figure \ref{fig:par-03} presents the light curve of this spectacular event, which exhibits long-term annual modulation due to the annual parallax effect. 
High brightness and the very accurately measured parallax allow us to constrain the mass and distance in a very tight relation, as shown in the bottom panel of Fig \ref{fig:par-03}.
This is the nearest possible mass-gap object in our sample at about 600 pc. 
The microlensing event had two solutions, one of which was rejected due to the potential contribution of blended light to the Gaia DR2 solution. 
Therefore, detailed follow-up observations are encouraged to verify the correctness of this assumption, for example with high angular resolution imaging in order to attribute the blended light either to the lens or to a nearby source of light. 

Among the remaining lenses there are other less massive candidates for mass-gap objects: 
PAR-04 with mass estimates of $2.9^{+1.4}_{-1.3}$ and $3.2^{+1.3}_{-1.3}$ $\msun$, 
PAR-12 with an estimated mass of $2.7^{+1.2}_{-1.2}$ $\msun$ (a new event, not in the Wyrz16 sample), 
PAR-39 with estimated masses of $2.6^{+1.7}_{-1.3}$ and $2.3^{+1.9}_{-1.0}$ $\msun$ for the two solutions, and 
PAR-34 with estimated masses of $2.0^{+1.2}_{-0.9}$ and $2.1^{+1.2}_{-0.9}$ $\msun$.
The masses of these lens candidates are  slightly higher with respect to the estimates from Wyrz16. 

The mass estimates for the three higher mass black holes in our sample have also increased, with 
PAR-02 at $11.9^{+4.9}_{-5.2}$ $\msun$, 
PAR-13 at $9.0^{+3.9}_{-3.7}$ and
$8.0^{+4.1}_{-3.2}$ $\msun$ for the two solutions, and
PAR-05 at $6.7^{+3.2}_{-2.7}$ $\msun$.
The estimated masses are higher with respect to the estimates in Wyrz16 because the proper motion of the source is now better constrained with the Gaia DR2 data, hence only part of the relative lens-source proper motion parameter space is favoured out of the broad prior region considered in Wyrz16. Specifically, GDR2 data support better determined, higher values of the relative proper motion $\murel$, and since the mass is proportional to the relative proper motion, we generally obtain more massive lenses with smaller error bars. 
The trend toward greater proper motion in Gaia data may indicate that black holes receive kicks at birth;  continued microlensing observations could be used not only to constrain black hole masses, but also to measure their natal kicks.

\section{Discussion}
In this work we utilised Gaia DR2 astrometric information on parallax microlensing events found in OGLE-III \citep{Wyrz16}. 
The source distances and proper motions from Gaia were included in the Bayesian estimate of the mass and distance of the dark lens. 
This improved the mass estimates for dark lens candidates from Wyrz16 and revealed new cases of likely compact-remnant lenses in that sample.
The new sample includes 10 of the dark lens candidates from the 13 in Wyrz16.
The PAR-08 event was lost because it had no Gaia DR2 information,
and the probabilities that the lens is a dark compact remnant dropped in PAR-19 and PAR-28 events given the new constrains from GDR2 (\ie the lens may be consistent with a main sequence star based on the blended light visible in the light curve).

On the other hand, our sample was expanded with eight new events. Four of them (PAR-12, -30, -38, -48) were not present in Wyrz16 because they were located outside of the Red Clump region on the CMD (see Fig.\ref{fig:cmd}).  
The remaining four events (PAR-03, -22, -24, -34), which previously were not classified in the dark lens sample,  had their dark lens probabilities increased to above the 75 \%\ threshold with the Gaia DR2 data, with PAR-03 reaching a 94.5 \%\ probability of hosting a compact-object lens. 
This is due to the fact that the sources in these events turned out to be closer than 8kpc (as assumed in Wyrz16), hence the lenses moved closer and for their mass there was insufficient blended light to explain the lens as a main sequence star. 

Our analysis included Gaia DR2 parameters such as the proper motions and the probability distribution on the distance computed following CBJ18. As emphasised in multiple Gaia DR2 publications (e.g., \citealt{Luri2018, vanLeeuwen2018DR2, GaiaDR2astro, GaiaDR2photo}) the distances for individual stars can only be used when the parallax error is relatively small and the parallax is large (nearby sources). For sources at a few kiloparsec with small parallaxes extra caution needs to be taken in using Gaia DR2 distances because of the complex error budget in the Gaia parallax measurements.
Following equation \ref{eq:mass}, a change in the distance to the source does not directly affect the measurement of the mass.
However, in order to obtain the posterior probability for the mass distribution of each event we compute the weights related to the density of stars in the Galaxy and their proper motions following \cite{Batista2011} and \cite{Skowron2011}. Since the weights depend on the distance to the lens ($\piL$), which is a function of $\piE$ and $\piS$, our weights will also depend on the assumed distance to the source; in other words, the estimated distance to the source impacts the range of possible lens distances, which in turn impacts the range of lens proper motions, and hence the range of relative source-lens velocities used for inference.
%

In Wyrz16 we derived the detection efficiency as a function of the event parameters, \ie how representative  a given event is for the underlying population of lenses with similar mass and distance. 
For the sample used here it is essentially not possible to derive a robust efficiency since it is convolved with Gaia detection efficiency, which in turn is very complex and depends on many factors, including selection criteria on the quality of astrometric parameters. 
Therefore, as an approximation, here we used the same efficiency function as in Wyrz16.
We assume that for bright events like ours, all located in the same direction (toward the Galactic bulge), Gaia's contribution to the efficiency is more or less flat across different timescales.
Since 41 sources out of 59 were matched, the detection efficiency of Gaia is around 70 \%\ and depends mostly on the magnitude and the crowding.


The distribution of medians shown in Figure \ref{fig:mass} is somewhat similar to that in Wyrz16 and does not visually display an obvious mass gap between neutron stars and black holes. The low-mass end with sub-solar mass white dwarfs is not present due to our selection of long-duration events with a detectable annual parallax (hence generally more massive) and with an obvious parallax effect (hence generally closer). This selection ignores short-timescale events due to stars, white dwarfs, and neutron stars at larger distances (beyond about 3 kpc), but also excludes events due to very massive lenses with large $\thetaE$ as their $\piE$ gets smaller for larger Einstein radii ($\piE=\pi_\mathrm{rel}/\thetaE$) and might not be detectable. On the other hand, for both short timescale and very massive lenses without a detectable microlensing parallax we would not be able to constrain the mass of the lens with just one parameter $\tE$. 

Bayesian hierarchical modelling of the underlying mass distribution of dark lenses based on the full posterior probability density functions of the 18 events does not find evidence for a broad mass gap between neutron star and black hole masses.  In our set of phenomenological models, we find that the observed mass distribution is best described as a single power law with slope $p(M) \propto M^{-3.2}$ start from a minimum mass of $M_\textrm{min}=2 M_\odot$.  The observed sample is also consistent with a mixture of a flat distribution of masses between 1 and 2 $M_\odot$ representing heavy white dwarfs and neutron stars and a power-law slope extending from $3 M_\odot$ representing black holes.  On the other hand, a mass gap between $2$ and $5$ $M_\odot$ is strongly disfavoured by the data within our range of models.

We therefore conclude that microlensing data support a continuum of masses, without an obvious mass gap.  This would indicate that the dynamical mass estimates of BH masses in BH X-ray binaries suffer from significant evolutionary selection effects that present as observational biases, as suggested by \citet[\eg][]{NarayanMcClintock2005}.  The absence of a mass gap would constrain the supernova explosion mechanism, 
favouring models that allow for significant accretion onto the proto-NS before an explosion can be driven by a growing instability, as in the `delayed' models of \citet{Fryer2012} and \citet{Belczynski2012}.



So far, we have assumed that compact remnants display the same peculiar velocity distribution relative to their host environments as regular stars.  However, neutron stars are observed to have large natal kicks \citep{Hobbs:2005}, and black holes may also receive natal kicks from asymmetric mass ejection during supernovae \citep[e.g.,][]{Willems:2005,Fragos:2009,Fryer2012,Repetto:2012, Repetto:2017, Gandhi2019} (but see \citealt{Mandel:2015kicks,Mirabel:2016}).

The angular radius of the Einstein ring $\thetaE$ is proportional to the square root of the lens mass, $\thetaE \propto \sqrt{M}$.  Meanwhile, the light curve timescale is set by the ratio of the Einstein radius and the relative angular velocity of the source and lens, $\tE = \thetaE / \murel$.  The physical transverse velocity depends on the angular velocity of the lens (obtained from $\murel$ by subtracting the angular velocity of the source, measured with Gaia) and the distance to the lens, on which $\thetaE$ also depends, creating a degeneracy between the lens velocity and the lens distance.  However, in general, there is a strong positive correlation between the lens velocity and the lens mass.  In other words, if the compact-remnant lenses have systematically higher velocities than their local stars, greater lens masses would be needed to explain these observations.

We can therefore ask how high the lens velocities would need to be in order to recreate the mass gap simply by forcing events with mass estimates in the putative mass gap to have higher masses of $5 M_\odot$.  
We find that rather modest lens velocities with the respect to the local  Galactic motion could be sufficient.
Excess peculiar velocities (presumably associated with natal kicks) of 20 km/s for PAR-03, 20 to 40 km/s for PAR-04, 40 km/s for PAR-12, 40 to 60 km/s for PAR-15, and 80 km/s for PAR-30 would be consistent with mass solutions of $5 M_\odot$ for these lenses, recreating the mass gap.  These are moderate velocities relative to the observed NS velocities, and are not unreasonable for low-mass black holes.   Therefore, this degeneracy between the lens mass and lens velocity represents an important caveat to our inference on the compact-remnant mass distribution. 
Conversely, high-mass BH candidates such as PAR-02 would require high natal velocities of $\gtrsim 100$ km/s to significantly affect the mass solutions.

All our microlensing events reported here have been modelled with a single-lens microlensing model and there were no signatures of any systematic deviations to it.
Given that in all events the maximum amplification is relatively high (\ie they have a small impact parameter $\u0$),  if the lens were a binary, we should expect to see the effects related to the caustic in the light curve \citep[\eg][]{Skowron2007, Shvartzvald2015}.
The separation in time between caustic entry and caustic exit amplification spikes in the light curve depends on the size of the caustic and the relative proper motion of the lens and the source. 
On the other hand, the OGLE-III typical light curve sampling is about 3 days, meaning we have sensitivity to caustics higher than 
$\frac{3}{365}*\murel$, which with a conservative assumption of $\murel=4$ mas/yr, gives the minimum caustic size of about 0.04 mas.  At a typical distance to our lenses of 1-2 kpc, this corresponds to about 0.04-0.08 AU. Since the size of the caustic reflects roughly the separation between the binary components, it means that in the OGLE-III events we should see the caustic effects for binaries wider than about 8-16 $R_\odot$. For binaries with closer orbits we are unable to distinguish them from single lenses and the mass we report is the sum of masses of the binary components. Therefore, it is still possible that some of our mass-gap objects are in fact binary neutron stars.
On the other hand, the formation rate of close double neutron star binaries in the Galaxy (see, e.g., \citealt{Tauris:2017,VignaGomez:2018}) is expected to be two orders of magnitude below the formation rate of black holes; therefore, double neutron stars are unlikely to be a source of significant contamination in our sample.

\section{Conclusions}
We re-investigated 59 parallax microlensing events found among 150 million stars observed in OGLE-III 2001-2009 data from \cite{Wyrz16} and used Gaia DR2 astrometric information to derive the probability they contain a dark lens.  
The new sample consists of 18 events where the light contribution from the lens cannot be explained with main sequence star and hence the lenses are most likely white dwarfs, neutron stars, or black holes.
The most likely masses of three of the lenses are above 5 $\msun$, and hence these are our strongest candidates for black holes,  the most massive of which (PAR-02) weighing in at $11.9^{+4.9}_{-5.2} M_\odot$. 
Additionally, we identified eight lenses with median masses between 2 and 5 $\msun$, \ie located in the putative `mass gap' observed in X-ray binary systems.
Hierarchical modelling of the underlying mass distribution behind our samples rules out a broad $2 M_\odot$ to $5 M_\odot$ mass gap between NS and BH masses. 
On the other hand, the masses of most `gap' systems would be consistent with falling above $5 M_\odot$ if they received natal kicks of $20$ to $80$ km/s. If we impose the existence of the mass gap as an a priori constraint, these measurements allow us to infer that natal kicks are likely to fall into this range for $\sim 5 M_\odot$ black holes.

Our work shows that gravitational microlensing can identify potential remnants and derive their masses if additional second-order effects are present (annual or space microlensing parallax).  
A measurement of the angular Einstein radius would break near-degeneracies and allow a unique determination of the mass of the lens as well as the velocity of the lens. This might be  possible with the brighter events observed by Gaia \citep{Rybicki2018}; however, it will be a challenge for fainter events. In the forthcoming era of the Large Synoptic Survey Telescope (LSST), which should detect hundreds of events similar to the ones studied here, sub-milliarcsecond astrometry will still be a challenge. Therefore, we anticipate that the method presented in this work will remain useful for estimating the masses of a larger population of compact-object lenses in near future.

\section*{Acknowledgments}
The authors would like to thank Drs. Chris Belczy{\'n}ski, Christopher Berry, Floor van Leeuwen, Zuzanna Kostrzewa-Rutkowska, Bernhard Mueller, Jan Skowron and the entire OGLE team for their encouragement and help at various stages of this work. 

{\L}.W. acknowledges support from the Polish NCN grants: Harmonia No. 2012/06/M/ST9/00172, Harmonia No. 2018/06/M/ST9/00311, Daina No. 2017/27/L/ST9/03221 and 
MNiSW grant DIR/WK/2018/12.  IM is a recipient of the Australian Research Council Future Fellowship FT190100574.

\bibliography{bibs}
\bibliographystyle{aa}

\end{document}